\documentclass[traditabstract]{aa} 

\usepackage{graphicx}
\usepackage{txfonts}
\usepackage{longtable}
\usepackage{pdfpages}

\begin{document}

\title{Stellar mass function of cluster galaxies at z $\sim$ 1.5: evidence for reduced quenching efficiency at high redshift}
\author{Julie B. Nantais\inst{1}, Remco F. J. van der Burg\inst{2}, Chris Lidman\inst{3}, Ricardo Demarco\inst{4}, Allison Noble\inst{5}, Gillian Wilson\inst{6}, Adam Muzzin\inst{7}, Ryan Foltz\inst{6}, Andrew DeGroot\inst{6}, and Mike Cooper\inst{8}}

\institute{Departamento de Ciencias F\'isicas, Universidad Andres Bello, Fernandez Concha 700, Las Condes 7591538, Santiago, Región Metropolitana, Chile
 \and
    Laboratoire AIM-Paris-Saclay, CEA/DSM-CNRS-Universit\'e Paris Diderot, Irfu/Service d'Astrophysique, CEA Saclay, Orme des Merisiers, 91191 Gif-sur-Yvette, France
 \and
   Australian Astronomical Observatory, PO Box 2915, North Ryde NSW 1670, Australia 
 \and
    Departamento de Astronom\'ia, Universidad de Concepci\'on, Casilla 160-C, Concepci\'on, Región del Biobío, Chile
 \and
    Department of Astronomy \& Astrophysics, University of Toronto, Toronto, Ontario M5S 3H4, Canada
 \and
    Department of Physics and Astronomy, University of California-Riverside, 900 University Avenue, Riverside, CA 92521, USA
 \and
    Institute of Astronomy, University of Cambridge, Madingley Road, Cambridge CB3 0HA, United Kingdom
 \and
    Department of Physics and Astronomy, University of California, Irvine, 4129 Frederick Reines Hall, Irvine, CA 92697, USA
}
   \date{Received ???; accepted ???}


  \abstract{We present the stellar mass functions (SMFs) of passive and star-forming galaxies with a limiting mass of 10$^{10.1}$ M$_{\odot}$ in four spectroscopically confirmed Spitzer Adaptation of the Red-sequence Cluster Survey (SpARCS) galaxy clusters at 1.37 $<$ z $<$ 1.63.  The clusters have 113 spectroscopically confirmed members combined, with 8-45 confirmed members each. We construct $Ks$-band-selected photometric catalogs for each cluster with an average of 11 photometric bands ranging from $u$ to 8 $\mu$m.  We compare our cluster galaxies to a field sample derived from a similar $Ks$-band-selected catalog in the UltraVISTA/COSMOS field.  The SMFs resemble those of the field, but with signs of environmental quenching.  We find that 30 $\pm$ 20\% of galaxies that would normally be forming stars in the field are quenched in the clusters.  The environmental quenching efficiency shows little dependence on projected cluster-centric distance out to $\sim$ 4 Mpc, providing tentative evidence of pre-processing and/or galactic conformity in this redshift range.  We also compile the available data on environmental quenching efficiencies from the literature, and find that the quenching efficiency in clusters and in groups appears to decline with increasing redshift in a manner consistent with previous results and expectations based on halo mass growth.}


\authorrunning{Nantais et al.}
\titlerunning{redshift 1.5 clusters}

\maketitle

\section{Introduction}

The study of galaxy clusters in formation at high redshift (z) is essential to our understanding of the relationship between structure formation and galaxy evolution in the Universe.  It is particularly important for understanding how the morphology-density relation (Dressler 1980) arose from the early z $>$ 2 protoclusters, which were just beginning to form their red sequences (Kodama et al.~2007; Tanaka et al.~2013).  Unlike lower redshift galaxy clusters, these protoclusters are usually rich in massive, star-forming galaxies (e.g., Overzier et al.~2008; Galametz et al.~2010; Hatch et al.~2011a, 2011b; Kuiper et al.~2010; Koyama et al.~2013a; Cooke et al.~2014; Dannerbauer et al.~2014; Shimakawa et al.~2014; Umehata et al.~2015).  However, the galaxies in protoclusters are still affected by their environment. The high-redshift protoclusters are known to show biases with respect to the field in terms of enhanced active galactic nucleus (AGN) activity in their central galaxies (Hatch et al.~2014), excesses of high-mass and/or dusty star-forming galaxies with respect to the field (Hatch et al.~2011b; Cooke et al.~2014), and accelerated chemical evolution (Shimakawa et al.~2015).  When compared to lower redshift clusters in terms of their halo masses and galaxy evolution, these protocluster environments show the expected differences in galaxy evolution from their cluster counterparts consistent with being the progenitors of the clusters (e.g., Shimakawa et al.~2014; Cooke et al.~2015).

At z $\sim$ 1, when many galaxy clusters are massive, mature, established, and resemble low-redshift clusters, environmental effects on typical cluster galaxies have shifted.  Quenching effects, morphological transformation from late-type to early-type, and rapid brightest-cluster-galaxy growth characterize these clusters.  The transformation from star-forming to passive in cluster galaxies at z $\sim$ 1 out-paces that of field environments and galaxy groups (van der Burg et al.~2013; Balogh et al.~2016), although the star formation rates of galaxies that are not yet quenched appear unaffected by environment (Muzzin et al.~2012; Koyama et al.~2013b).  The transformation from star-forming to passive is more advanced in galaxies whose positions and/or radial velocities suggest they have been in the galaxy cluster for a longer period of time (Muzzin et al.~2012, 2014; Nantais et al.~2013a,b; Noble et al.~2013, 2016).  The quenched fractions and early-type fractions of cluster galaxies are independently correlated with both stellar mass and environment (Muzzin et al.~2012; Nantais et al.~2013a), and the state of galaxy evolution in z $\sim$ 1 clusters generally appears to be independent of the selection method (X-ray or infrared) of the galaxy clusters as well (Foltz et al.~2015).  In addition to z $\sim$ 1 being an epoch of enhanced quenching of cluster galaxies, it is also an epoch of rapid brightest-cluster galaxy growth which parallels that of the cluster halos in general (Lidman et al.~2012, 2013).

In recent years, the expanding area of research on galaxy clusters and protoclusters in between the above two epochs, at 1.3 $<$ z $<$ 2, has shown this intermediate epoch to be an essential phase in galaxy cluster growth.  In this redshift range, the role of environment appears to be transitioning between protocluster environmental effects (e.g., growth of massive galaxies via dusty star formation) and cluster effects (e.g., quenching of lower-mass galaxies).  Galaxy clusters recognized in and around this epoch often still show substantial star formation in their central regions, unlike clusters at lower redshifts (e.g., Brodwin et al.~2013; Fassbender et al.~2014; Bayliss et al.~2014; Webb et al.~2015a,b).  However, environmental quenching is starting to be very important, since many are starting to show substantial red sequences and excesses of quenched galaxies compared to the field, even slightly before this epoch (e.g., Kodama et al.~2007, Bauer et al.~2011; Quadri et al.~2012; Gobat et al.~2013; Strazzullo et al.~2013; Andreon et al.~2014; Newman et al.~2014; Balogh et al.~2016; Cooke et al.~2016).  Brightest-cluster galaxy growth is also important at these redshifts, with evidence for both gas-rich mergers (Webb et al.~2015a,b) and gas-poor mergers (Lidman et al.~2012, 2013) contributing to this growth.  There are also indications that galaxy evolution correlates with local environment within a forming cluster in this redshift range, as is found at z $\sim$ 1 (Hatch et al.~2016).  In order to truly complete our understanding of the evolution of environmental effects on galaxies from protoclusters to mature clusters, we need to study large and, to the extent possible, homogeneously-observed samples of growing high-redshift galaxy clusters at various epochs, rather than a few individual systems.

The Spitzer Adaptation of the Red-Sequence Cluster Survey (SpARCS; Wilson et al.~2009; Muzzin et al.~2009) has been an excellent resource in high-redshift galaxy cluster research, with more than a dozen spectroscopically confirmed, infrared-selected galaxy clusters already reported in the literature (Demarco et al.~2010; Muzzin et al.~2012; Muzzin et al.~2013a; Lidman et al.~2012, 2013; Webb et al.~2015a,b).  Recently-processed data from this survey allow us to expand our knowledge of the critical phases of galaxy cluster evolution with several z $>$ 1.3 galaxy clusters observed and selected in a manner similar to their lower redshift counterparts.  In this paper, we analyze the stellar mass functions and environmental quenching efficiencies in a sample of four SpARCS galaxy clusters in the range 1.37 $\leq$ z $\leq$ 1.63.  The clusters have 113 spectroscopic members between them, and two of the clusters are described for the first time.  We compare these clusters with the state-of-the-art UltraVISTA/COSMOS field sample (Muzzin et al.~2013b) in order to discern effects attributable to the cluster environment.  In Section 2 we describe the photometric and spectroscopic data.  In Section 3 we describe the methods of photometric and spectroscopic analysis we used to build our catalogs.  In Section 4 we describe how we obtained our stellar mass functions.  In Section 5 we provide the results, and in Sections 6 and 7 we provide a discussion and summary.  In this paper, we assume a $\Lambda$CDM cosmology with H$_0$ = 70 km s$^{-1}$ Mpc$^{-1}$, $\Omega_M$ = 0.3, and $\Omega_{\Lambda}$ = 0.7.  All magnitudes used in this paper are in the AB system.

\section{Data}

Our galaxy clusters were identified using the Stellar Bump Sequence (SBS) technique described in detail in Muzzin et al.~(2013a).  This infrared selection technique uses the 1.6 $\mu$m stellar bump, a feature prominent in the rest-frame near-infrared spectral energy distribution of virtually any galaxy containing an underlying stellar population of at least intermediate age ($\geq$ 100 Myr).

At z $\sim$ 1.5, the 1.6 $\mu$m stellar bump conveniently spans the {\it{Spitzer}} Infrared Array Camera (IRAC) 3.6 and 4.5 $\mu$m bands, such that galaxies red in [3.6]$-$[4.5] are likely to be at high redshifts (Papovich 2008).  Since galaxies at 0.2 $<$ z $<$ 0.4 are also quite red in [3.6]$-$[4.5], a color cut in $z'-[3.6]$ is made to screen out foreground galaxies at z $<$ 0.8 (Muzzin et al.~2013a).  Using this two-color system, several high-redshift cluster candidates were identified within SpARCS and later spectroscopically confirmed.  

The spectroscopic data used to confirm the southern clusters and the photometric data used to obtain photometric redshifts, stellar masses, and rest-frame colors are described in the following subsections.  A brief summary of the photometric and spectroscopic data for each system, plus the number of confirmed members, is shown in Table 1.

We show $gz[3.6]$ images of these clusters in Figure 1, in the same style as their lower redshift counterparts in Wilson et al.~(2009) and Muzzin et al.~(2009).  Objects in white boxes are spectroscopically confirmed cluster members, the bulk of which are small and red (or purple) in these images. Some bright red objects near the centers of clusters lack spectroscopic redshifts despite being targeted in our observing runs, since at z $>$ 1.3 it is difficult to obtain the signal-to-noise needed for absorption-line redshifts. Photometric redshifts of these bright red objects show that most have a high probability of being passive cluster members.  The two lower redshift clusters in the top panels of Figure 1, SpARCS-J0335 (z = 1.369, 22 spectroscopic members) and SpARCS-J0225 (z = 1.598, 8 spectroscopic members), are being described and analyzed for the first time.  The highest-redshift cluster, SpARCS-J0224 (z = 1.633, 45 spectroscopic members), was featured in Muzzin et al.~(2013a).  SpARCS-J0224 also appeared in Lidman et al.~(2012) along with SpARCS-J0330 (z = 1.626, 38 spectroscopic members).  Here we extend the analysis of these clusters beyond the few bright cluster members and photometric merger candidates identified in those papers.

\begin{figure*}
\centering
\includegraphics[width=16cm]{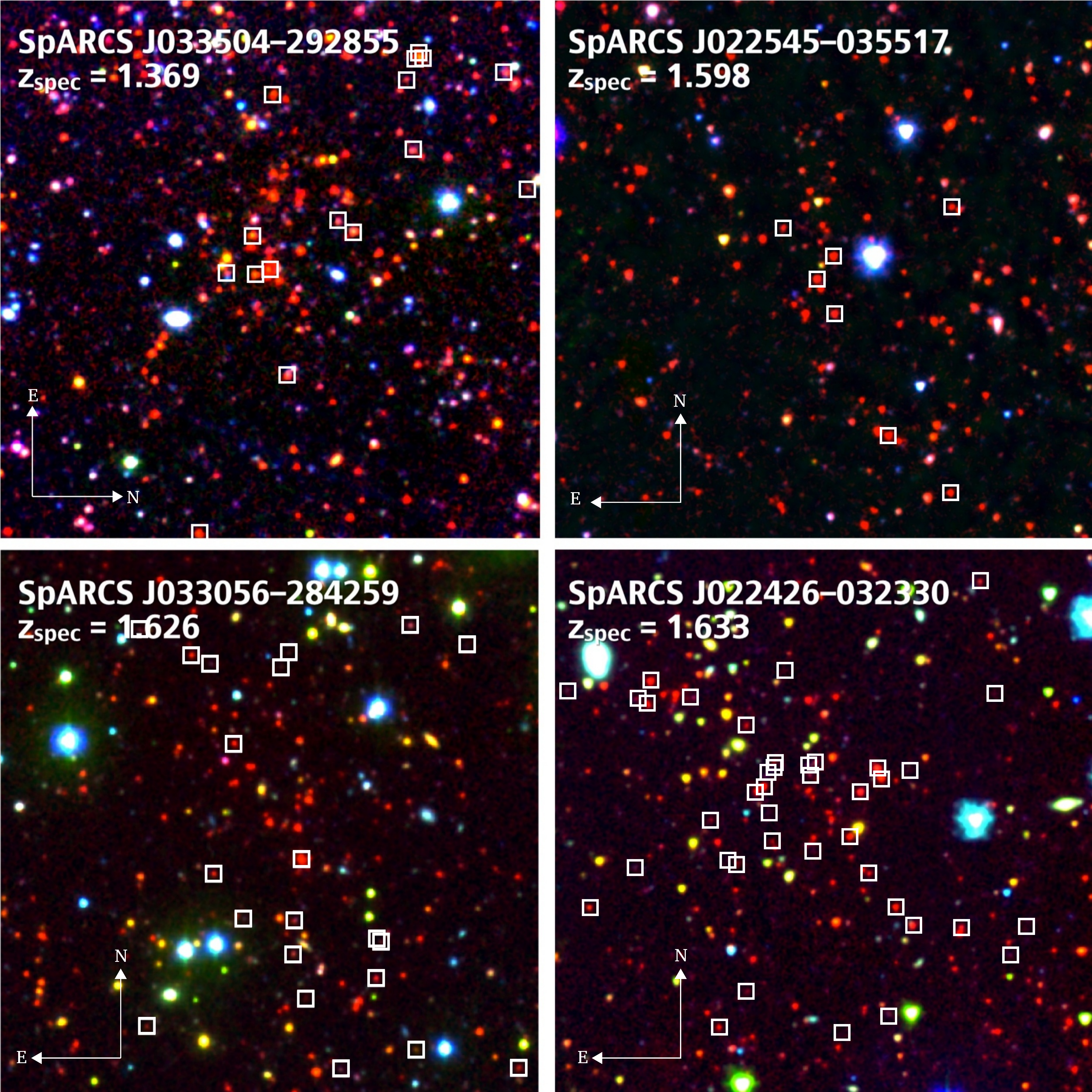}
\caption{Tri-color $gz[3.6]$ images of the central regions of the four southern z $>$ 1.35 SpARCS clusters, with 4$\arcmin$ $\times$ 4$\arcmin$ fields of view.  Spectroscopic cluster members are marked with white squares, and the orientation of the image is shown in the lower left corner of each cluster image.}
\label{images}
\end{figure*}

\begin{table*}
\caption{SpARCS high-redshift data summary}
\label{tab1}
\centering
\begin{tabular}{lcccccc}
\hline\hline
Cluster & RA (J2000) & Dec (J2000) & z & Spectroscopy & Photometry & Spec. members \\ 
 & h:m:s & d:m:s & & & &  \\ 
\hline
SpARCS-J0224 & 02:24:26.33 & -03:23:30.8 & 1.633 & FORS2, MOSFIRE, OzDES & $ugrizYJKs$3.6$\mu$m4.5$\mu$m5.8$\mu$m8.0$\mu$m & 45 \\
SpARCS-J0330 & 03:30:55.87 & -28:42:59.5 & 1.626 & FORS2, MOSFIRE, OzDES & $ugrizYJKs$3.6$\mu$m4.5$\mu$m5.8$\mu$m8.0$\mu$m & 38 \\
SpARCS-J0225 & 02:25:45.55 & -03:55:17.1 & 1.598 & FORS2, MOSFIRE, OzDES & $ugrizYKs$3.6$\mu$m4.5$\mu$m5.8$\mu$m8.0$\mu$m & 8 \\
SpARCS-J0335 & 03:35:03.58 & -29:28:55.6 & 1.369 & FORS2, OzDES & $grizYKs$3.6$\mu$m4.5$\mu$m5.8$\mu$m8.0$\mu$m & 22 \\
\hline  
\end{tabular}
\end{table*}

\subsection{Spectroscopy}
Spectroscopy of high-redshift cluster member candidates was performed with the Focal Reduction and Imaging Spectrograph 2 (FORS2; Appenzeller \& Rupprecht 1992) on the European Southern Observatory (ESO) Very Large Telescope (VLT) Unit Telescope 1 (UT1) in Mask Exchange Unit (MXU) mode for all four clusters.  The spectroscopy was performed in four separate service-mode programs.  Total on-target integration time ranged from 40 minutes to four hours.  Data reduction was performed with the customized software described in Nantais et al.~(2013b) for FORS2 MXU data.

For SpARCS-J0330, SpARCS-J0224, and SpARCS-J0225, near-infrared multi-object spectroscopy was obtained with the MOSFIRE spectrograph on the Keck Telescopes in Hawaii in several observing runs (PI G.~Wilson).  Comparison of MOSFIRE vs. FORS2 redshifts for overlapping objects in SpARCS-J0224 and SpARCS-J0330 suggests an uncertainty of +/- 0.0013 in our spectroscopic redshifts.  This uncertainty is similar to that found in Nantais et al.~(2013b) for FORS2 spectroscopy of a z = 1.2 galaxy cluster, and lower than the velocity dispersion expected for galaxy clusters at these redshifts.

Additional spectroscopy exists for all fields in the Australian Dark Energy Survey (OzDES; Yuan et al.~2015).  OzDES targets typically lie at lower redshift than these clusters.  However, one redshift is of a cluster member, a bright AGN in the outskirts of SpARCS-J0335.  The remainder of the OzDES redshifts, at z $<$ 1, are very useful for checking our photometric redshifts and improving their quality.

Table 1 shows the number of spectroscopic members in each cluster (totaling to 113), along with a brief summary of the data.  A substantial concentration of foreground galaxies, likely a sheet or filament, with $\sim$ 11 members at z $\sim$ 1.4 was also discovered in the SpARCS-J0225 field, but we do not analyze this system.

\subsection{Imaging}

The imaging data were collected with ground- and space-based facilities and cover a wavelength range that, for most clusters, extends from the observer-frame $u$ band (0.35 $\mu$m) to 24 $\mu$m (with {\it{Spitzer}} MIPS).  In our analysis, we use the data out to IRAC 8 $\mu$m. Three of the four clusters were observed in the optical with the f/2 camera of the Inamori-Magellan Areal Camera \& Spectrograph (IMACS; Dressler et al. 2011) on the Magellan Baade telescope during two observing runs, one in September 2011 and the other in December 2012. The fourth cluster, SpARCS J0225, lands within the deep D1 field observed by the Canada-France-Hawaii Telescope (CFHT) Legacy Survey\footnote{We used version 7 of the reduced images, which are available from http://terapix.iap.fr} (CFHTLS) and therefore has deep coverage in the optical with MegaCam on CFHT. All four clusters were observed in the near-infrared with the High Acuity Wide-field K band Imager (HAWK-I; Pirard et al. 2004; Casali et al. 2006) on the ESO VLT in service mode over three ESO periods.  

The clusters were also imaged with IRAC on the {\it{Spitzer}} Space Telescope, first as part of the Spitzer Wide-area Infrared Extragalactic Survey (SWIRE; Lonsdale et al.~2003) in all four IRAC bands, and then later with deeper observations in the IRAC 3.6 and 4.5 $\mu$m bands as part of the Spitzer Extragalactic Representative Volume Survey (SERVS).

\subsubsection{IMACS and CFHTLS data processing}

The IMACS data were processed in a standard manner using our own scripts 
that called IRAF\footnote{IRAF is distributed by the National Optical Astronomy 
Observatories which are operated by the Association of Universities
for Research in Astronomy, Inc., under the cooperative agreement
with the National Science Foundation} tasks. 

SCAMP (Bertin 2006) and SWarp (Bertin et al.~2002)\footnote{http://www.astromatic.net/} were used to map the sky-subtracted $z$-band images onto the astrometric reference frame defined by bright stars in the USNO-B1 catalog.  The internal residuals reported by SCAMP were typically 0.03$\arcsec$ or less. We used these $z$-band images as the astrometric reference
for all other images taken with IMACS and HAWK-I. The uncertainties in the zero points vary from 1\% in $i$ to 5\% in $u$.  For SpARCS-J0225, we used the
CFHTLS $i$-band image as the astrometric reference due to incomplete areal coverage in the $z$ band.

\subsubsection{HAWK-I data processing}

The processing of the raw HAWK-I data was also done in a standard manner and largely follows the steps outlined in Lidman et al.~(2008, 2012). SCAMP and SWarp were used to map the sky-subtracted images onto the astrometric reference frame defined by the bright stars in the $z$-band images that were taken with IMACS. After accounting for irregularities such as gain variations, bad pixels, and satellite trails, the images were then combined with the IMCOMBINE task within IRAF. Each image was weighted with the inverse square of the full width at half maximum (FWHM) of the point spread function (PSF) to maximize the image quality of the final combined image.

With the exception of the data taken in the $Y$ band, zeropoints were set using stars from the the Two-Micron All-Sky Survey (2MASS) point source catalog (Skrutskie et al. 2006). For $Y$, the zeropoint was set using standard stars that were observed during the
same night as the clusters. The uncertainties in the zeropoints are generally less than 2\%, and more typically 1\%, for both $J$ and $Ks$. 

Before performing multi-color photometry, all processed optical, HAWK-I, and IRAC data were transformed to a common pixel scale using SWARP to facilitate photometry.  The final scale of all images was 0.185$\arcsec$ pixel$^{-1}$.

\subsubsection{Data quality, including image depth}

Over all, the image quality of the ground-based imaging data varies from
0.3$\arcsec$ FWHM for $Ks$-band data taken with HAWK-I to 1.1$\arcsec$ FWHM for $u$-band IMACS data.

$Ks$-band detection completeness limits were calculated using a methodology similar to that used for UltraVISTA in Muzzin et al.~(2013b).  The ten brightest non-saturated PSF stars in each field were used to make a model PSF in the $Ks$-band images (transformed to the 0.185$\arcsec$ pixel$^{-1}$ scale, but without adjustments of image quality) using the IRAF DAOPHOT package.  This PSF was then dimmed to create artificial point sources.  Two thousand copies of these point sources were superimposed upon the $Ks$-band detection images in random locations.  An artificial point source was considered detected if (a) a detection was made within one pixel (0.185$\arcsec$) of its true location and (b) this detection did not coincide with a preexisting source in the $Ks$-band detection images.  Our method corresponds to the realistic scenario of Muzzin et al.~(2013b), accounting for losses due to overlap with bright stars and poor-quality regions of the image.  With it, we estimate a maximum completeness of 99\% at $Ks$ $<$ 21 mag AB.

\begin{figure}
\centering
\includegraphics[width=9cm]{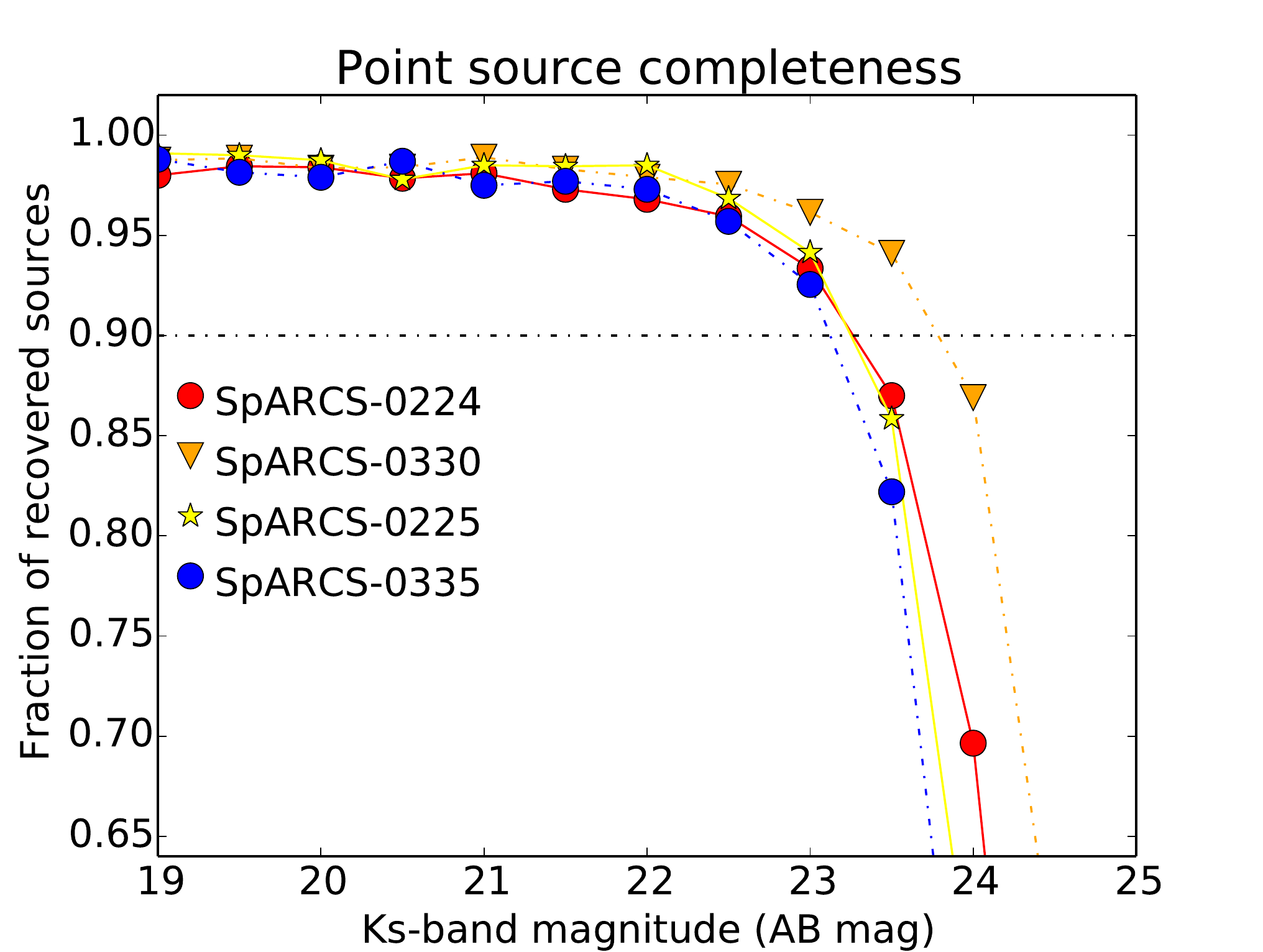}
\caption{$Ks$-band detection completeness fraction as a function of total $Ks$ magnitude for the four southern cluster fields.  The red circles represent SpARCS-J0224; the orange triangles represent SpARCS-J0330; the yellow stars represent SpARCS-J0225; and the blue circles represent SpARCS-J0335.  The black dash-dot line represents the 90\% cutoff for determining the completeness limit.}
\label{completeness}
\end{figure}

The resulting completeness curves for each field are shown in Figure 2. The $Ks$-band completeness limits are shown in Table 2, which is discussed in Section 3.2 along with the corresponding stellar-mass completeness limits.  Although the empirical completeness limit for SpARCS-J0330 is 23.78 mag, we adopt a value of 23.40 mag in Table 2 due to the lack of UltraVISTA field comparison objects fainter than this magnitude.  We use the 90\% completeness limits for consistency with van der Burg et al.~(2013).

\section{Photometric and spectroscopic analysis}

\subsection{Photometric catalog creation}
The photometric catalogs were created following procedures similar to those of Muzzin et al.~(2013b) for UltraVISTA.  Objects were detected in the same $Ks$-band images as were used for the completeness using Source Extractor (Bertin \& Arnouts 1996) version 2.19.5.  Astrometric and pixel-scale matching was performed on all images using SWarp prior to photometry.  

To determine colors, PSF matching was performed with convolution kernels created with the IRAF task {\it{wiener}}\footnote{The task {\it{wiener}} applies a non-iterative Fourier deconvolution filter chosen from one of four varieties: inverse, Wiener, geometric mean, or parametric.  The default fitting parameters with a Wiener filter, recommended by the authors of the task for restoring stellar images, were used to calculate the convolution kernels from our stars.} on all images.  The PSFs of optical and ground-based near-infrared bands were matched to the poorest image quality among these bands.    

Photometry for color determination was measured in aperture diameters of 2.2$\arcsec$ in the $u$ through $Ks$ bands and 3.7$\arcsec$ in the IRAC bands using Source Extractor in dual-image mode.  Correction factors for $Ks$-IRAC colors were calculated in a manner similar to Muzzin et al.~(2013b) and van der Burg et al.~(2013).

All photometry intended for science is corrected for Galactic extinction (reddening) using Schlegel, Finkbeiner \& Davis (1998) dust maps and Schlafly and Finkbeiner (2011) extinction corrections.  Extinction values were very low in the infrared, less than 0.01 mag in the $Ks$ band and IRAC, and low to moderate in the optical: 0.02-0.05 mag in the $i$ band and 0.03-0.09 mag in the $g$ band.

We estimated the photometric uncertainties via the fluctuations of background levels in up to 5000 empty apertures, similar to the methodology of Labb\'e et al.~(2003), with non-empty apertures rejected at the extremes of the flux distribution.  Additional uncertainties of 5\% of the flux for IRAC and 1\% of the flux for optical through $Ks$ were added to account for zeropoint uncertainties.

The zeropoints for the multi-wavelength aperture photometry were tested and, if need be, corrected using the stellar locus (High et al.~2009) as compared to Covey et al.~(2007) and stars in the GCLASS catalogs.  The adjustments (1-10\%) roughly matched the uncertainties (up to a few percent) in most bands, but were sometimes larger in $J$ and $Ks$. These larger offsets were still comparable to those of Muzzin et al.~(2013b), in which $Ks$-band corrections were about 8\%. Integrated photometry in $Ks$-band was determined using Source Extractor's MAG\_AUTO parameter, with an aperture correction (typically around 4\%) determined using artificial stars of known flux.

\subsection{Photometric redshift and rest-frame color estimation}
Photometric redshifts and rest-frame $UVJ$ colors were calculated using the EAZY code (Brammer, van Dokkum \& Coppi 2008).  Stellar masses were estimated using the FAST code (Kriek et al.~2009).  Parameters for EAZY and FAST were based on those used in van der Burg et al.~(2013), and the EAZY parameters in particular were chosen to minimize the scatter in photometric redshifts.  FAST stellar mass outputs were based on Bruzual \& Charlot (2003) [BC03] models.

\begin{figure}
\centering
\includegraphics[width=9cm]{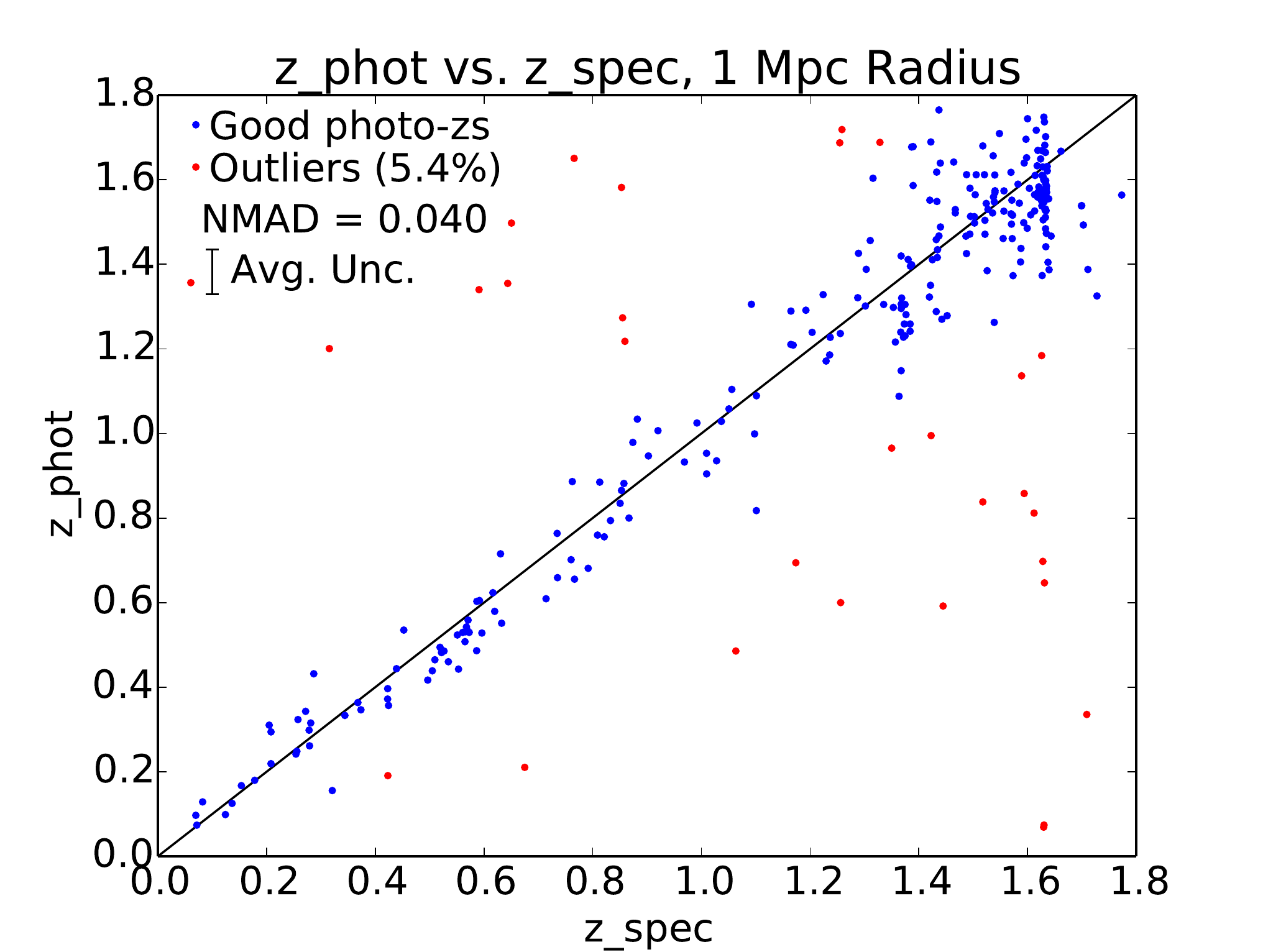}
\caption{Photometric vs.~spectroscopic redshift for the inner 1 Mpc sample of galaxies in the four cluster fields.  The normalized median absolute deviation (NMAD) and outlier rates are included in the figure.}
\label{RedshiftBig}
\end{figure}

Figure 3 shows the photometric vs.~spectroscopic redshifts for galaxies up to 1 Mpc from the cluster center in all four cluster fields.  Redshifts significantly beyond those of the highest-redshift clusters are virtually absent, given that the spectral features needed to confirm redshifts fall outside our FORS2 and MOSFIRE windows at z $\sim$ 1.7.  The scatter (normalized median absolute deviation or NMAD) of $(z_{phot}-z_{spec})/(1+z_{spec})$ (excluding outliers) is $\sigma$ = 0.04, comparable to van der Burg et al.~(2013). Our outlier rate within this region, defined as percentage of objects with $(z_{phot}-z_{spec})/(1+z_{spec})$ $\geq$ 0.15 as in van der Burg et al.~(2013), is 5.4\%, slightly higher than in van der Burg et al.~(2013).  Outlier rates without cluster-centric distance constraints increase by about a factor of 3 due to inclusion of more faint objects, but this has little effect on typical photometric redshift quality. Redshifts above those of the highest-redshift clusters are nearly absent since the spectral features needed to confirm redshift fall out of the optical and near-IR windows of our spectroscopic data.

The stellar masses derived from FAST allowed us to determine stellar-mass completeness limits corresponding to our $Ks$-band completeness limits.  Following the example of van der Burg et al.~(2013), we determine these stellar-mass limits as being the highest stellar mass of any UltraVISTA galaxy at our $Ks$-band completeness limit near the cluster redshift.  In Table 2 we show the stellar mass limits determined for each cluster, featured in Column 4.

\begin{table*}
\caption{Stellar mass completeness limits}
\label{tab2}
\centering
\begin{tabular}{lcccc}
\hline\hline
Cluster & Redshift & $Ks$ limit & Log mass limit & Mass function limit\\ 
 & & mag & log(M/M$_{\odot}$) & log(M/M$_{\odot}$) \\ 
\hline
SpARCS-J0224 & 1.633 & 23.26 & 10.39 & 10.50 \\
SpARCS-J0330 & 1.626 & 23.40 & 10.30 & 10.30 \\
SpARCS-J0225 & 1.598 & 23.25 & 10.33 & 10.50 \\
SpARCS-J0335 & 1.369 & 23.12 & 10.10 & 10.10 \\
\hline  
\end{tabular}
\end{table*}

\subsection{Photometric members and comparison samples}
We define photometric membership for our clusters as  $(z_{phot}-z_{cl})/(1+z_{cl})$ $\leq$ 0.05, following van der Burg et al.~(2013), after adjusting the photometric redshifts by by $(z_{phot}-z_{cl})/(1+z_{cl})$ = +0.02 to account for a slight systematic offset.  Our photometric membership cut-off closely matches the scatter of our photometric redshifts ($\sigma$ $\sim$ 0.04).  We exclude objects near bright stars from our photometric member sample.

The UltraVISTA field galaxies were used as comparison samples for the clusters.  UltraVISTA field objects were selected if their photometric redshifts and stellar masses were within the same range as photometric cluster members and they did not overlap with bright stars.

\begin{figure}
\centering
\includegraphics[width=9cm]{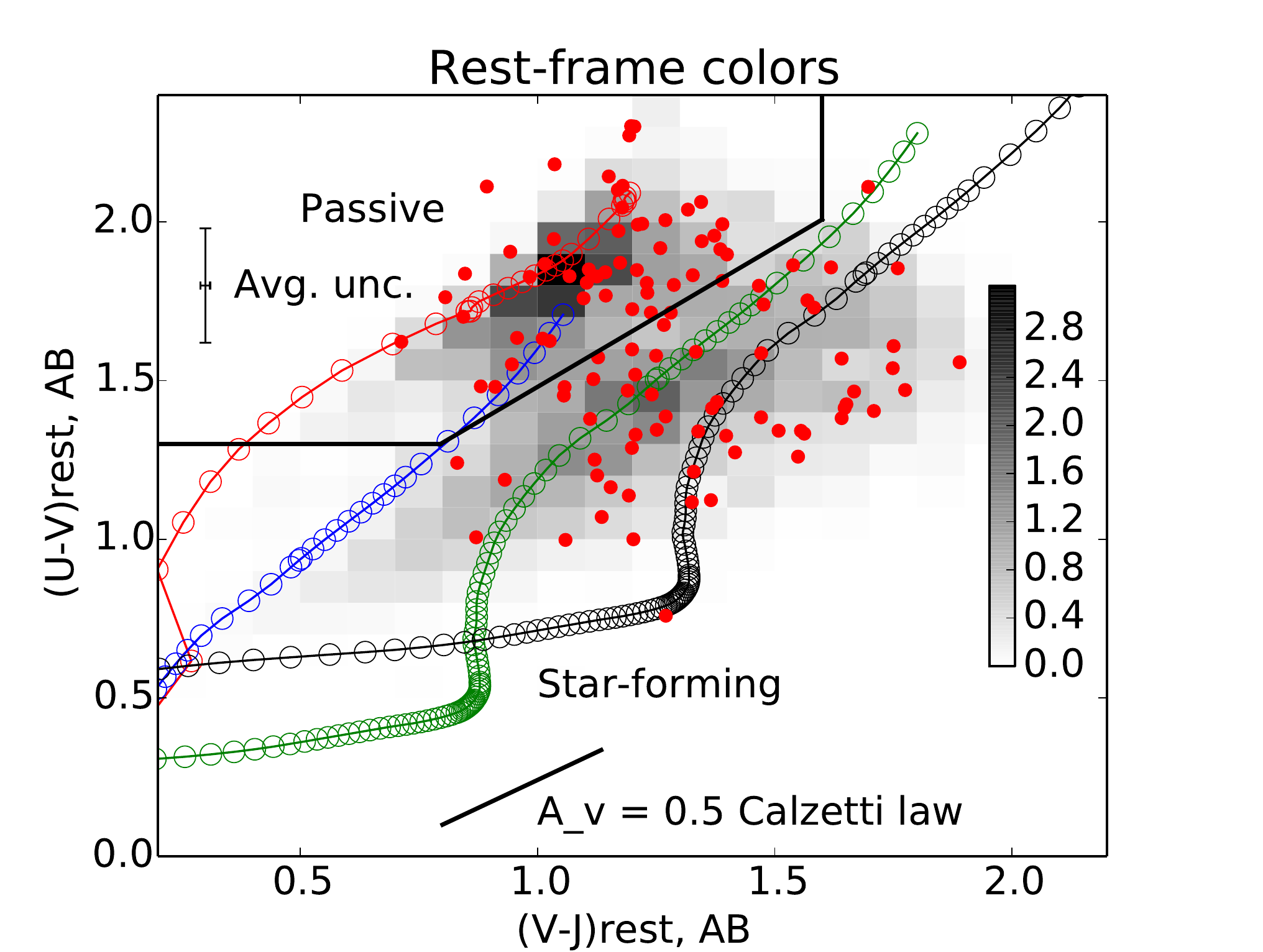}
\caption{Rest-frame $UVJ$ colors for photometric and spectroscopic cluster members (red dots), shown with the color distribution for UltraVISTA field galaxies of the same stellar mass and photometric redshift as the clusters (gray scale in units of arbitrarily normalized UltraVISTA galaxy counts).  Solar-metallicity Bruzual \& Charlot (2003) complex stellar population evolutionary tracks are displayed with variable age steps (15,000 yr to 0.25 Gyr) out to 4.5 Gyr (which corresponds to the age of universe at z = 1.37).  The red model represents a single burst and the blue, green, and black models represent exponentially declining star formation with increasing levels of internal reddening.}
\label{uvjdiagram}
\end{figure}

The rest-frame $UVJ$ color-color diagram for photometric and spectroscopic cluster members and UltraVISTA field galaxies, which indicates the passive vs.~star-forming distinction (Wuyts et al.~2007, Williams et al.~2009, Patel et al.~2012), is shown in Figure 4.  The cluster galaxies (confirmed and unconfirmed) are shown as red dots, while the arbitrarily normalized, stellar-mass-matched UltraVISTA field galaxy counts are shown in gray scale.  The rest-frame $UVJ$ colors of the cluster galaxies appear to be generally well-matched to the UltraVISTA field galaxy distribution, suggesting similar accuracy in the color determination.  The passive fractions are also similar, though slightly higher in the cluster sample: 45\% of the spectroscopic and photometric cluster members are passive vs. 39\% of the field galaxies.

\section{Determination of stellar mass functions}

To determine stellar mass functions, we followed a methodology based on van der Burg et al.~(2013).  We defined seven stellar mass bins starting at 10$^{10.1}$ M$_{\odot}$ separated by 0.2 dex, and estimated the total number counts (spectroscopic members plus photometric members minus field counts) in each bin within 1 Mpc of the cluster center.  In each cluster, the cluster center is defined as the position of the brightest (in $Ks$) galaxy located in the region with the highest surface density of photometric and spectroscopic members.  Spectroscopic member counts are given a Poisson uncertainty, while photometric member counts are given an uncertainty based on 100 Monte Carlo simulations of photometry varied within its error margins.  In the two lowest stellar mass bins, the total counts are corrected for the fraction of clusters complete in these bins.

Because of the bias in our spectroscopic selection, in which false positives for photometric membership outweigh false negatives due to easy confirmation of bright red star-forming foreground galaxies, the spectroscopy-based field contamination correction factors of van der Burg et al.~(2013) are inappropriate for our data.  We therefore corrected the photometric cluster members for field contamination by estimating field galaxy counts within $(z_{phot}-z_{cl})/(1+z_{cl})$ $\leq$ 0.05 of the cluster center and subtracting these counts, yielding a correction independent of the quality of spectroscopy.  We use the full UltraVISTA data to correct for the estimated field galaxy counts in each cluster, scaling the total counts to match the survey volume for each cluster.  

We correct for field counts separately for passive and star-forming galaxies by subtracting the passive galaxy field counts from the passive stellar mass function and the star-forming field counts from the star-forming stellar mass function.  Passive and star-forming galaxies are each divided into three V-J color bins and added together in order to make the total passive and star-forming stellar mass functions, and all color bins for both passive and star-forming galaxies are added together to make the total stellar mass functions.  With our methodology, we obtain the same total stellar mass functions regardless of whether or not we treat the passive and star-forming galaxies separately.  Similarly, the division of passive and star-forming galaxies into multiple color bins gives the same overall passive and star-forming stellar mass functions as would treating all passive and all star-forming galaxies as a single entity.

Uncertainties in field counts are estimated by resampling the UltraVISTA field in the areas and redshift ranges of all four clusters (equivalent to 4$\pi$ Mpc$^2$ in projected area) in 1000 random locations, and taking the standard deviation of the total field counts (for all four clusters) in these 1000 samplings as the sampling uncertainty in the field corrections.  Again, this is done separately for passive and star-forming galaxies as well as for the full sample combined.  The uncertainties from field sampling are similar in value to the field counts themselves, and are added in quadrature to the cluster galaxy count errors.  The field sampling uncertainty contributes moderately to the error margins in the stellar mass functions and the passive and star-forming fractions, and strongly to the uncertainty in the environmental quenching efficiency (discussed in Section 5.2).

In the high mass bins, we include all galaxies, including the brightest cluster galaxies (BCGs) of the clusters, unlike van der Burg et al~(2013).  This is due to a lack of evidence that the BCGs in our particular high-redshift clusters deviate from the Schechter (1976) function, unlike those in massive lower redshift clusters.  In fact, in all four clusters, the $Ks$-band BCG is either not the most massive galaxy or is tied for the most massive with a fainter galaxy.  In contrast, only 3 of the 10 similarly analyzed z $<$ 1.35 GCLASS clusters have a BCG that is not also the most massive cluster galaxy according to our team's unpublished data. Therefore, the evolutionary processes that would create extraordinary BCGs that deviate from the Schechter function have not yet occurred in our high-redshift clusters.

The seven stellar mass bins with total photometric and spectroscopic cluster member counts and uncertainties are fitted with a Schechter function using a maximum-likelihood method. Since we do not account for asymmetrical error bars in the number counts, this fit is basically equivalent to least-squares fitting.  However, in determining uncertainties in the stellar mass function itself, we do choose to take into account the asymmetries in the error bars.  We use emcee (Foreman-Mackey et al.~2012) to perform Monte Carlo simulations of various fluctuations around the maximum-likelihood parameters $\phi^*$ (normalization), log ($M^*$) (characteristic mass), and $\alpha$ (faint-end slope).  We take the 1$\sigma$ uncertainties to be the difference between the best-fit values and the lower and upper limits (marginalizing over other parameters) of each parameter for all simulated results with $(ln(L_{max}) - ln(L))$ $<$ 0.5, following van der Burg et al.~(2013), where $L$ is the likelihood function for the given set of fit parameters.

The UltraVISTA stellar mass functions are estimated as in van der Burg et al.~(2013) as the number of galaxies per stellar mass bin per co-moving cubic megaparsec, with Poisson uncertainties.  The survey volume was estimated to be 1.09 $\times$ 10$^7$ Mpc$^3$ in the redshift range 1.251 $<$ z $<$ 1.765, which corresponds to $|(z_{phot}-z_{cl})/(1+z_{cl})| \leq 0.05$ of the lowest- and highest-redshift clusters.  In the lowest stellar mass bin, only the SpARCS-J0335 redshift range was considered, and the counts in this bin were corrected for the fraction of survey volume coverage.

\section{Results}

Below we discuss the stellar mass functions and their uncertainties for the clusters and the field, and estimate the environmental quenching efficiency of the clusters.

\subsection{Stellar mass functions}
The stellar mass functions for the total, passive, and star-forming galaxy populations are shown in Figure 5, with the cluster stellar mass functions on the left and the field stellar mass functions on the right.  The van der Burg et al.~(2013) stellar mass function fits and type fractions, normalized to match our data, are shown as dash-dot lines.  Type fractions (passive in red, star-forming in blue) based on rest-frame $UVJ$ colors are shown below the stellar mass functions in Figure 5, with uncertainties in type fractions for the clusters derived from the same Monte Carlo simulations as were used to estimate the uncertainties in galaxy counts.  

The shape of the total stellar mass function is virtually identical between the cluster and field, within 1.1$\sigma$.  In the field, the passive component is dominant at high stellar masses, while the star-forming component clearly dominates at low masses.  In the clusters, however, passive galaxies make up roughly 50\% of the total even at low masses, as opposed to only 20\% at low masses in the field (29\% in total), indicating the importance of environmental quenching.  Throughout most of the range, the passive and star-forming fractions remain within 1$\sigma$ of one another in the clusters.  In the highest-mass bin there is only a single star-forming galaxy, an unconfirmed photometric member, and therefore its presence does not allow us to draw any definitive conclusions about the high-mass end of the stellar mass function.  In the Schechter function fit, its main effect is to increase the uncertainty in M$^*$.

When comparing to the van der Burg et al.~(2013) curves (dash-dot lines), both the cluster and the field show quenching between z $\sim$ 1 and z $\sim$ 1.5.  The quenching is much more drastic in the clusters, where passive galaxies become heavily dominant at nearly all stellar masses at the lower redshifts.  The total mass functions, however, change very little between z $\sim$ 1.5 and z $\sim$ 1, apart from a modest increase in high-mass galaxies seen in the field.

Table 3 summarizes the results of the parameter fits for log M$^*$ and $\alpha$ and gives a reduced $\chi$$^2$ goodness-of-fit (GoF) estimate.\footnote{The large GoF values in the field are driven by the small fractional uncertainties in galaxy counts (Poisson or scaled Poisson) as compared to their deviations from the Schechter function, such that the Schechter function appears to under-fit the data.  The small GoF values in the clusters are driven by large fractional uncertainties in galaxy counts due to small number counts and large uncertainties from photometry, such that the Schechter function appears to over-fit the data.}  All uncertainties in Table 3 are based on the lower and upper limits from Monte Carlo simulations of parameter sets in which $(ln(L_{max}) - ln(L))$ $<=$ 0.5.  The M$^*$ and $\alpha$ values agree within 1$\sigma$ between the clusters and the field.  Both the clusters and the field have shallow $\alpha$ for the passive galaxies and similar, steeper $\alpha$ for star-forming galaxies.  The values in the table, like the figures, generally support the notion of similar stellar mass functions for both the cluster and the field, for star-forming and passive galaxies alike.

\begin{table}
\caption{Stellar mass function parameters}
\label{tab3}
\centering
\begin{tabular}{lcccc}
\hline\hline
Galaxy type & Environment & log(M*/M$_\odot$) & $\alpha$ & GoF \\ 
\hline
\\
All & Clusters & $10.70^{+0.23}_{-0.23}$ & $-0.48^{+0.81}_{-0.60}$ & 0.17 \\
\\
All & Field & $10.77^{+0.02}_{-0.02}$ & $-1.03^{+0.04}_{-0.05}$ & 21.05 \\
\\
Passive & Clusters & $10.58^{+0.27}_{-0.29}$ & $-0.07^{+1.35}_{-0.85}$ & 0.64 \\
\\
Passive & Field & $10.69^{+0.03}_{-0.02}$ & $-0.26^{+0.08}_{-0.08}$ & 10.72 \\
\\
Star-forming & Clusters & $10.94^{+0.77}_{-0.39}$ & $-1.06^{+1.02}_{-0.79}$ & 0.23 \\
\\
Star-forming & Field & $10.71^{+0.03}_{-0.02}$ & $-1.21^{+0.06}_{-0.06}$ & 10.72 \\
\\
\hline  
\end{tabular}
\end{table}

\subsection{Cluster quenching efficiency}

We estimate the total corrected passive and star-forming counts from the binned data and use these in the environmental quenching efficiency formula\footnote{We do not employ the van der Burg et al.~(2013) method of fitting the passive fraction to a superposition of field passive and star-forming stellar mass functions because the uncertainties were too large with our data.} employed by various other researchers (van den Bosch et al.~2008, Peng et al.~2012, Philips et al.~2014a, Kawinwanichakij et al.~2016, Balogh et al.~2016).  The environmental quenching efficiency, also called conversion fraction or environmentally-quenched fraction, is traditionally defined as follows:

\begin{figure*}
\centering
\includegraphics[width=18cm]{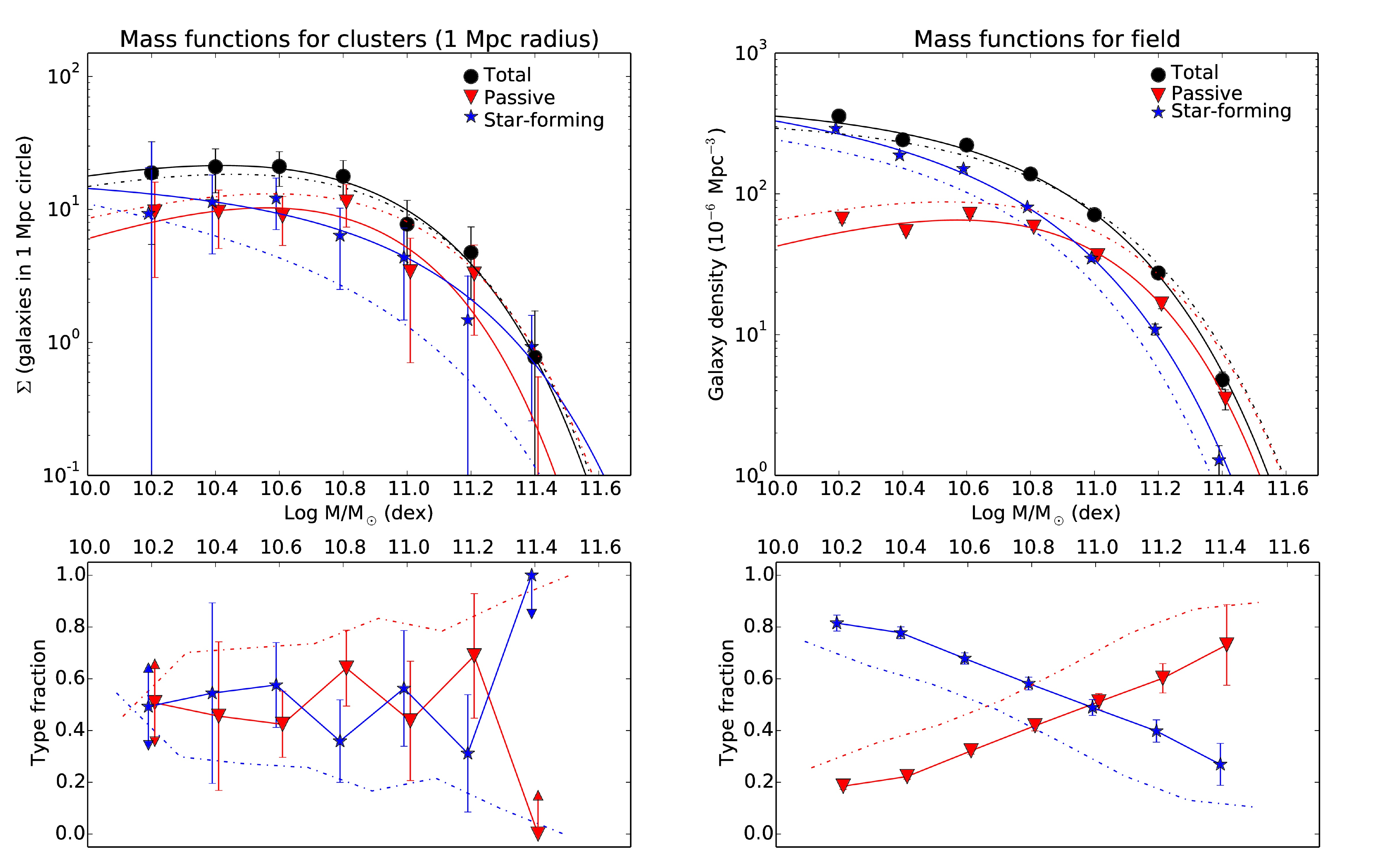}
\caption{Stellar mass functions for cluster (left) and field (right) galaxies, showing both the data and the best Schechter function fits (top) and the type fractions in each mass bin (bottom) with red representing passive galaxies and blue representing star-forming galaxies.  The dash-dot lines represent stellar mass functions and type fractions from van der Burg et al.~(2013) normalized to match our data.  Arrows for error bars represent unconstrained type fractions in a stellar mass bin.  Passive and star-forming stellar mass data are offset horizontally by $\pm$ 0.01 dex for clarity.}
\label{mass_func}
\end{figure*}

\begin{equation}
f_{eq} = (f_{passive, dense}-f_{passive, field})/(1-f_{passive, field})  
\end{equation}

In this equation, $f_{eq}$ is the environmentally-quenched fraction, $f_{passive, dense}$ is the passive fraction in the dense environment (cluster, protocluster, group, or filament), and $f_{passive, field}$ is the passive fraction in the field.  In our data, we obtain an environmental quenching efficiency of 30 $\pm$ 20\%, similar to Quadri et al.~(2012) for their z $\sim$ 1.6 cluster.

Our result for the environmental quenching efficiency is higher than (but consistent with) the typical 20\% environmental quenching efficiencies around passive central galaxies at z $\sim$ 1.5 in Kawinwanichakij et al.~(2016), but lower than (yet still consistent with) the environmental quenching efficiency of 45\% obtained at z $\sim$ 1  in van der Burg et al.~(2013) and the quenching efficiencies of 40\% and 60\% found by Balogh et al.~(2016) for groups and clusters at z $\sim$ 1 respectively.  Our results therefore suggest environmental quenching effects intermediate between rich clusters at z $\sim$ 1 and groups at z $\sim$ 1.5, although the uncertainties (driven largely by the field subtraction) do not allow statistical confirmation of this possibility.

Cooke et al.~(2016) do not calculate an environmental quenching efficiency directly for their z $\sim$ 1.6 high-redshift cluster, but among all galaxies above 10$^{10}$ M$_{\odot}$, 70 $\pm$ 13\% were deemed to be quenched in their high-redshift cluster, vs.~only 28\% in the UDS control field.  This would suggest an environmental quenching efficiency of 58\%, higher than the van der Burg et al.~(2013) value and comparable to the rich cluster value in Balogh et al.~(2016).  However, this is based on only 29 $\pm$ 6 excess galaxies, with very few spectroscopic members.  Given the quoted uncertainty in the original quenched fraction, the uncertainty of the Cooke et al.~(2016) environmental quenching efficiency would make it consistent with the results of van der Burg et al.~(2013).

\section{Discussion}

\subsection{Robustness of results}

In order to test the robustness of our results, we considered three important aspects of our sampling that might affect our stellar mass functions: (1) the areal extent of the cluster sample; (2) the redshift range of confirmed members; and (3) the correction for estimated field galaxy counts.

Although we consider only the inner megaparsec for consistency with van der Burg et al.~(2013), we may be excluding a large number of infalling galaxies which may be undergoing environmental pre-processing.  According to Muldrew, Hatch, and Cooke (2015), a network of galaxies spanning 50 co-moving megaparsecs in diameter at z $\sim$ 2 may be destined to become part of a z = 0 cluster.  At z = 1.6, the upper limit would correspond to a radius of about 9 (proper) Mpc for a massive cluster, which is larger than the area we probe.  We therefore check whether the inclusion of more distant infalling cluster members (both photometric and spectroscopic) would change our results.

We redid the analysis without any cluster-centric distance cuts (up to 4 Mpc from the cluster center).  Our results are shown in the left panels of Figure 6, with the 1 Mpc fits normalized to the full-survey counts and the 1 Mpc type fractions shown as dash-dot lines.  The contribution from high-mass galaxies is much higher within 1 Mpc than in the full survey, indicating that galaxies in the cluster outskirts are predominantly faint. The passive and star-forming fractions, however, are essentially unchanged.  The full survey yields an environmental quenching efficiency of 27\% $\pm$ 10\%, almost identical to our previous result (with smaller percent error due to much higher low-mass galaxy counts and a lower percent uncertainty from field subtraction).

To double-check the quenching efficiency results at large projected distances, we estimated the environmental quenching efficiency for all galaxies farther than 1 Mpc from the cluster centers.  We still found a value of 25 $\pm$ 16\% for the cluster outskirts when correcting for field galaxies (though only 10 $\pm$ 4\% without such corrections).  We rule out the possibility that the SpARCS-J0225 foreground sheet alone is mimicking pre-processing in the cluster outskirts, since excluding this cluster actually increases the environmental quenching efficiency of the cluster outskirts.

The above result has several possible explanations: (a) pre-processing of the low-mass galaxies in the cluster outskirts, (b) general galactic conformity (Weinmann et al.~2006; Kauffmann et al.~2013; Kauffmann 2015; Hearin, Watson, \& van den Bosch 2015), (c) rapid quenching after having crossed the cluster even once (Wetzel et al.~2013), or (d) difficulty distinguishing photometrically between passive and star-forming galaxies at the faint magnitudes typical of the cluster outskirts.  We note the representative error bar in U-V color in Figure 4 for galaxies within 1 Mpc, which includes nearly all the bright galaxies.  Regardless of the reason for this, however, including galaxies at larger cluster-centric distances does not affect our result, and thus we have no reason to discount our results on the basis of excluding distant, star-forming cluster members.

We also tested our results with a smaller sample of galaxies within the central 0.5 Mpc, in case the original 1 Mpc limit included too many non-members and recent infalls to show the full extent of environmental effects. We obtained nearly the same environmental quenching efficiency (31\% $\pm$ 26\%) as within a 1 Mpc radius.

\begin{figure*}
\centering
\includegraphics[width=18cm]{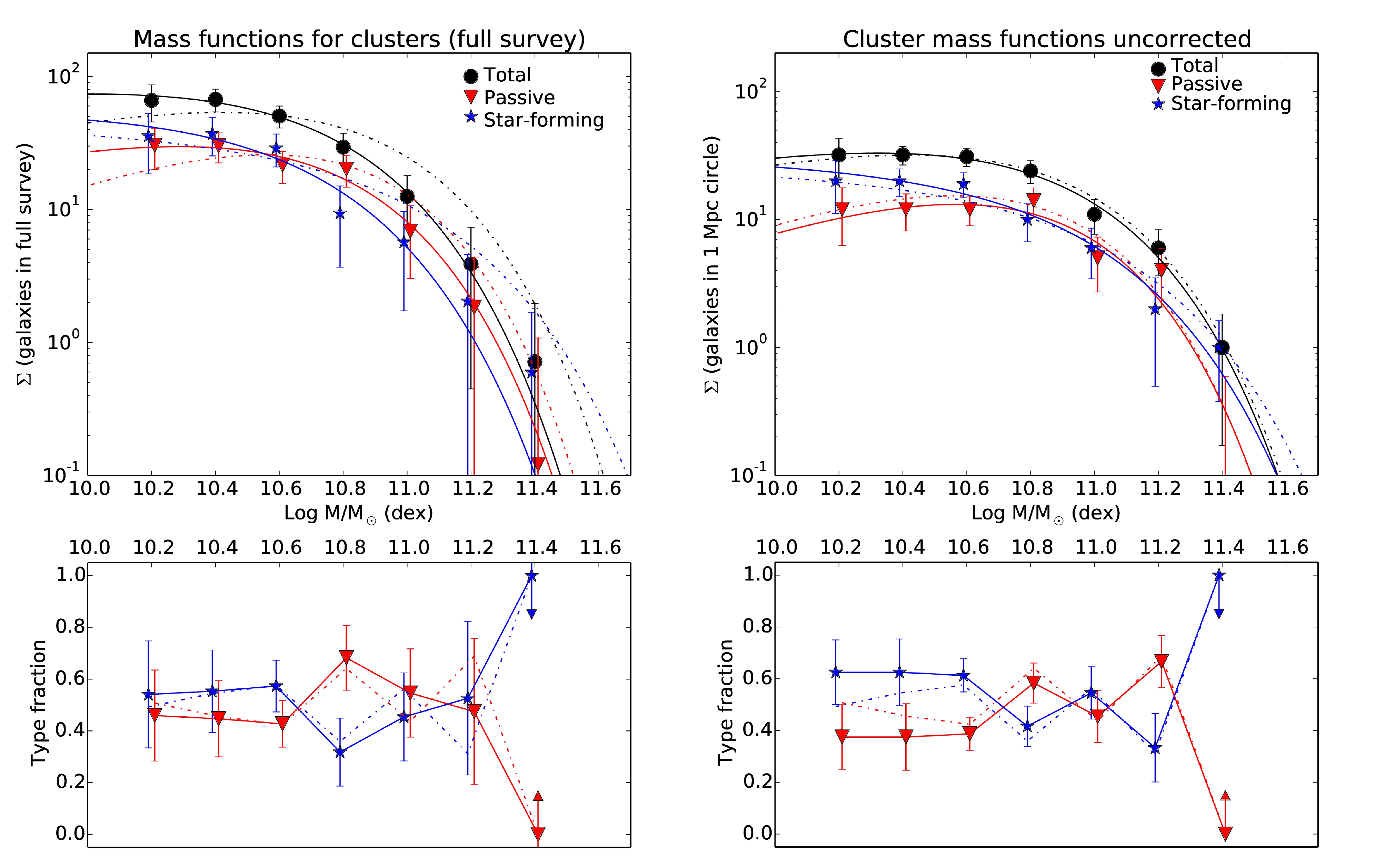}
\caption{Similar to Figure 5, but with the left panels showing the cluster stellar mass functions removing cluster-centric distance constraints and the right panels showing the cluster stellar mass functions in the inner 1 Mpc without field correction.  The dash-dot lines represent our original results normalized to match the reanalysis.}
\label{mass_func_4mpc}
\end{figure*}

\begin{figure*}
\centering
\includegraphics[width=18cm]{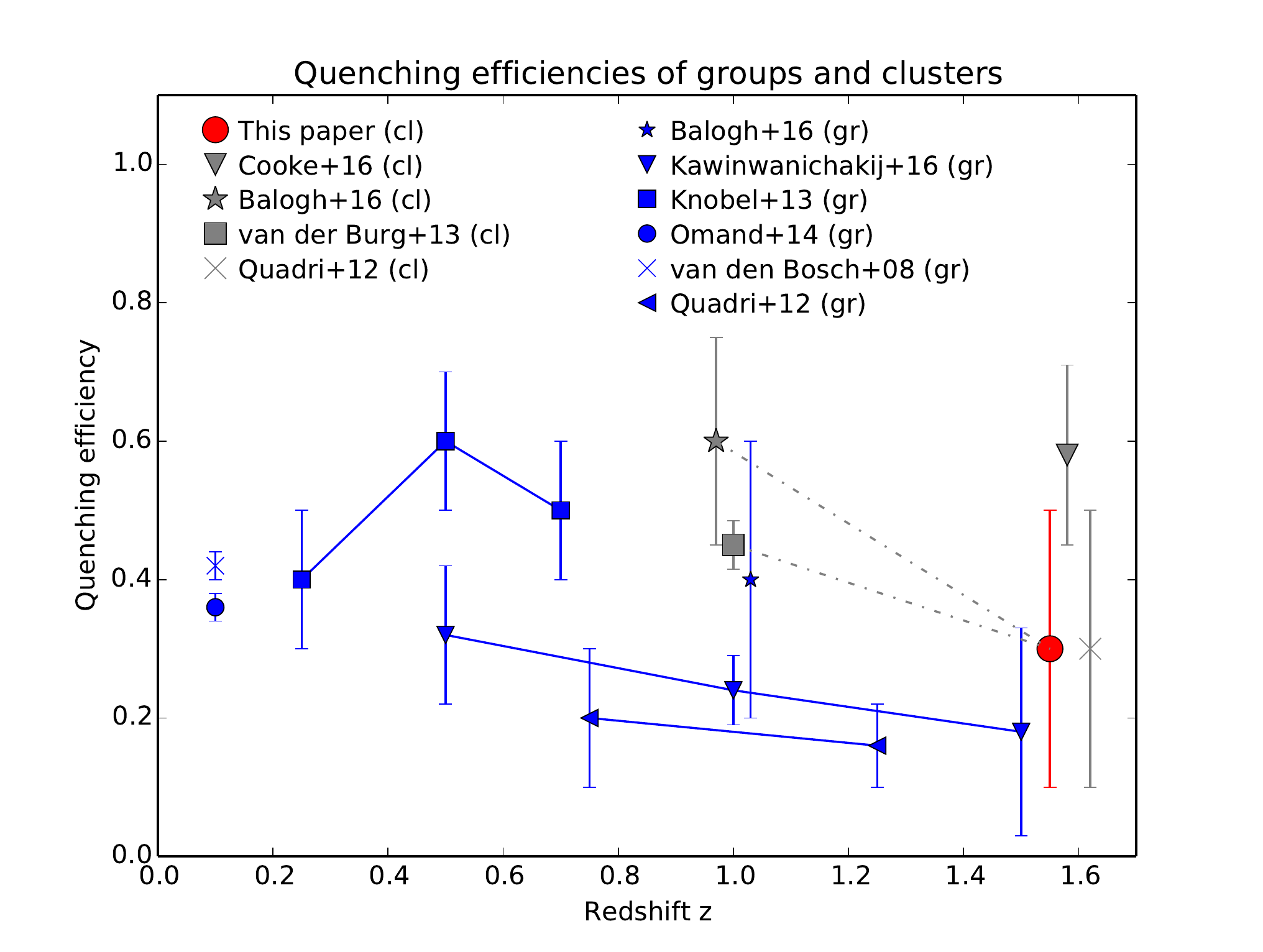}
\caption{Environmental quenching efficiencies as a function of redshift for our high-redshift cluster sample (large red circle) and various other group and cluster samples in the literature.  Dash-dot lines connect related studies and solid lines connect results from the same study.  Slight offsets between studies at the same redshift with overlapping error bars have been added for clarity.}
\label{quench}
\end{figure*}

Our next test addressed our spectroscopic member redshift cutoff. The original redshift cutoffs correspond to 3$\sigma$ of a rich cluster's velocity dispersion, which could be considered overly generous for a low-mass, non-virialized cluster (discounting complex superstructure along the line-of-sight).  We re-analyze the data with redshift cutoffs of $\pm$ 0.0075 from the cluster center at z $\sim$ 1.6 and $\pm$ 0.008 at z = 1.37, half the previous values.  These cutoffs still contain 50-87\% of the original spectroscopic members of each cluster, but span a maximum line-of-sight diameter of 18-19 Mpc.  The narrower redshift cutoff has no significant effect on mass function shapes or environmental quenching efficiency, raising the latter insignificantly to 31 $\pm$ 22\%.

Another check on the robustness of our results is to not attempt to subtract field galaxies from the photometric member counts --- that is, assume that every object that meets our photometric selection requirements is a cluster member.  This would lead to increased contamination from pristine field galaxies, thereby making our results less distinguishable from the field.  Finding signs of quenching even without field-galaxy corrections would therefore increase our confidence in our basic result.

The right panels of Figure 6 show the 1 Mpc cluster sample with no corrections for field contamination, with our original result shown as dash-dot lines.  The result is similar to the original but with less similarity in type fraction between passive and star-forming cluster galaxies at low masses. The environmental quenching efficiency in the uncorrected sample is 20\% $\pm$ 6\%, matching the Kawinwanichakij et al.~(2016) value for groups around passive galaxies at z $\sim$ 1.5.  If we repeat this no-field-correction analysis in the central 0.5 Mpc, the environmental quenching efficiency for the uncorrected sample rises to 26\% $\pm$ 8\%, approaching the value of $\sim$ 30\% consistently found after correcting for field galaxy counts.  The quenching efficiency values without field subtraction have much higher confidence (3$\sigma$ as opposed to 1.5$\sigma$) due to not including the large uncertainties from field counts. Therefore, we have high confidence in our main conclusions that (a) the cluster stellar mass functions are generally similar to the field, but with evidence of increased quenching relative to the field at low masses and (b) the environmental quenching efficiency of our clusters is lower than what is found in clusters at z $\sim$ 1 but comparable to or higher than that of groups at z $\sim$ 1.5.

We leave for the Appendix a final check on the robustness of our results: the comparison of field galaxies in UltraVISTA with the most pristine field galaxies of our own survey.

\subsection{Environmental quenching efficiencies and evolution of dense environments}

The intuitive picture of galaxy evolution that we would expect is that galaxies in clusters, protoclusters, and groups are quenched more efficiently over time, so that environmental quenching efficiencies at low redshifts would exceed those at high redshifts in a given category of local density or halo mass.  Gerke et al.~(2007) found some evidence of this when comparing the blue fractions of group galaxies at 0.75 $<$ z $<$ 1.3.  The blue fractions in the groups were lower than in the field at z $\leq$ 1, but by z $\sim$ 1.3 they were identical to field values.  Butcher \& Oemler (1984) and more recent studies such as Haines et al.~(2013) have found similar effects in rich galaxy clusters at low-intermediate redshifts.  

In Figure 7 we compare the environmental quenching efficiency of our high-redshift cluster sample with various environmental quenching efficiencies estimated from other studies of groups and clusters at various redshifts, with their references given in the legend.  The group samples are represented by blue symbols and the cluster samples are represented by gray symbols.  Our sample is represented by the large red circle.  

The quenching efficiencies appear to vary mostly by halo mass category (groups vs.~clusters), but within each halo mass category, there are signs of a decrease in environmental quenching efficiency with increasing redshift, with groups and clusters at z $\sim$ 1.5 having quenching efficiencies about 10\% lower than their similarly-selected counterparts at z $\lesssim$ 1.  Such a trend is consistent with the earlier findings of Gerke et al.~(2007) and Haines et al.~(2013).

Although increased quenching efficiencies and higher halo masses are both typical at lower redshifts, halo mass growth alone probably cannot account for the redshift evolution of quenching efficiency.  Various studies have found that environmental quenching at a given redshift does not have an exceptionally strong dependence on halo mass and is virtually independent of the stellar mass of the galaxy (Wetzel et al.~2013; De Lucia et al.~2014; Hirschmann et al.~2014; Phillips et al.~2014b; Fillingham et al.~2015; Wheeler et al.~2015).  Furthermore, interstellar medium research suggests that higher redshift galaxies use up their gas very quickly (e.g., PHIBSS: Tacconi et al. 2013; Genzel et al. 2012, 2015; Saintonge et al. 2013), an effect that may enhance quenching efficiency at high redshift if the gas supply is suddenly shut off.  Therefore, some other factor, such as the average amount of time spent in the cluster environment per galaxy, may lead to reduced environmental quenching efficiencies at high redshift.

When interpreting Figure 7 we note that the environmental quenching efficiency does not take into account the estimated epoch of infall of the galaxies in a cluster, and the evolution of the quenched fraction in the field --- its lower value at higher redshift --- since that epoch.  Therefore, environmental quenching efficiencies at lower redshifts may be underestimates of the true impact of the environment, if the galaxies in groups and clusters at low redshifts were accreted at higher redshifts and thus more likely to be star-forming before entering the cluster.  However, it is also possible that the galaxies that were accreted into clusters were pre-processed in groups beforehand, and thus were more quenched than typical field galaxies at their infall time, balancing this effect (McGee et al.~2009).

\section{Summary}

We analyzed the stellar mass functions of passive and star-forming galaxies down to 10$^{10.1}$ M$_{\odot}$ in a set of four spectroscopically confirmed, infrared-selected galaxy clusters from SpARCS at 1.37 $<$ z $<$ 1.63, with a total of 113 spectroscopic members and an extensive catalog of photometric redshifts, stellar masses, and rest-frame colors from an average of 11 optical to infrared bands.  Our major results were as follows:

The shape of the total stellar mass function of the galaxy clusters resembles that of the field.  The shapes of the passive and star-forming stellar mass functions also resemble those of the field, although there is less difference between their relative dominance at both high and low masses in the galaxy clusters.  That is, in the galaxy clusters the passive and star-forming fractions are close to 50\% at all stellar masses, whereas in the field passive galaxies are dominant at the highest stellar masses and star-forming galaxies are highly dominant at low stellar masses.  This difference between clusters and field is more significant at low stellar masses.  The environmental quenching efficiency, or conversion fraction, of our clusters was estimated to be 30 $\pm$ 20\%, intermediate between high-redshift groups at z $\sim$ 1.5 and galaxy clusters at z $\sim$ 1.

The above results were found to be robust with respect to cluster-centric distance, spectroscopic redshift cuts for cluster membership, and corrections for field contamination.  The robustness of type fractions and conversion fraction with respect to cluster-centric distance may be a signature of galactic conformity or group and filament pre-processing.  Not correcting for field contamination led to a modest decrease in the environmental quenching efficiency, but it differed significantly from zero and had a value comparable to high-redshift groups, which increased our confidence in our results.

A compilation of environmental quenching efficiency data from the literature indicates a decrease in environmental quenching efficiency in groups and clusters with increasing redshift, with a drop of about 10\% between z $\lesssim$ 1 and z $\sim$ 1.5.  This finding is consistent with previous research at intermediate redshifts finding less difference between groups and clusters and the field at higher redshifts than at lower ones.  Since related research suggests that galaxies should quench faster at higher redshifts (Balogh et al.~2016) and quenching is not strongly related to halo mass within a given redshift range, this result may be due to the galaxies in higher redshift groups and clusters, having spent, on average, significantly less time in the cluster environment than in lower redshift clusters.

\begin{acknowledgements}
Based in part on observations obtained with MegaPrime/MegaCam, a joint project of CFHT and CEA/DAPNIA, at the Canada-France-Hawaii Telescope (CFHT), which is operated by the National Research Council (NRC) of Canada, the Institut National des Sciences de l'Univers of the Centre National de la Recherche Scientifique (CNRS) of France, and the University of Hawaii. This work is based in part on data products produced at TERAPIX and the Canadian Astronomy Data Centre as part of the Canada-France-Hawaii Telescope Legacy Survey, a collaborative project of NRC and CNRS.

Based in part on observations taken at the ESO Paranal Observatory (ESO programs 085.A-0166, 085.A-0613, 086.A-0398, 087.A-0145, 087.A-0483, 088.A-0639, 089.A-0125, and 091.A-0478).

Based in part on observations taken at the Las Campinas Observatory in Chile.

Based in part on observations taken at the Keck Telescopes in Hawaii.

Based in part on observations made with the Spitzer Space Telescope, which is operated by the Jet Propulsion Laboratory, California Institute of Technology under a contract with NASA.

During the course of this project, JN received funding from FONDECYT postdoctoral research grant no.~3120233 and Universidad Andres Bello internal research project DI-651-15/R.

This research has made use of the VizieR catalog access tool, CDS, Strasbourg, France.

RD gratefully acknowledges the support provided by the BASAL Center for Astrophysics and Associated Technologies (CATA), and by FONDECYT grant no. 1130528.

RFJvdB acknowledges support from the European Research Council under FP7 grant number 340519.

\end{acknowledgements}

\begin{appendix}
\section{SpARCS vs.~UltraVISTA field galaxies}

As a final check on the robustness of our results, we compare the stellar mass functions of SpARCS and UltraVISTA field galaxies to look for any important systematic differences.  For the SpARCS data, we define field galaxies as objects with 1.251 $<$ z $<$ 1.487 (SpARCS-J0335 photometric redshifts) in the SpARCS-J0330 and SpARCS-J0224 fields.  This avoids cluster-outskirts members within SpARCS-J0330 and SpARCS-J0224.  We exclude SpARCS-J0225 from the field galaxy analysis because any foreground sample may include members of the z $=$ 1.43 sheet which might differ from pristine field members.  We also exclude SpARCS-J0335 due to the shallow optical photometry for this cluster making it impossible to distinguish between passive and star-forming z $\sim$ 1.6 background galaxies.  In the UltraVISTA field, we also consider only objects with 1.251 $<$ z $<$ 1.487 to compare to our SpARCS field galaxies, to ensure that the co-moving line-of-sight distance and state of galaxy evolution in the field are equivalent in both samples.

To construct the stellar mass function for SpARCS field galaxies, we take the average counts in each stellar mass bin in our 100 Monte Carlo simulations with photometric redshift and stellar mass results varied according to uncertainties in the photometry.  Averaging the simulated results, as opposed to using the original counts only, helps reduce variations due to small-number statistics, except in the two highest mass bins in which the SpARCS foreground field samples are most deficient.  These highest mass bins would be expected to be occupied largely by brightest group and cluster galaxies, whereas the SpARCS field subsample was selected to be free of known groups and clusters.  Stellar mass functions were fit the same way as for the clusters and for the field in the main analysis.

\begin{figure*}
\centering
\includegraphics[width=18cm]{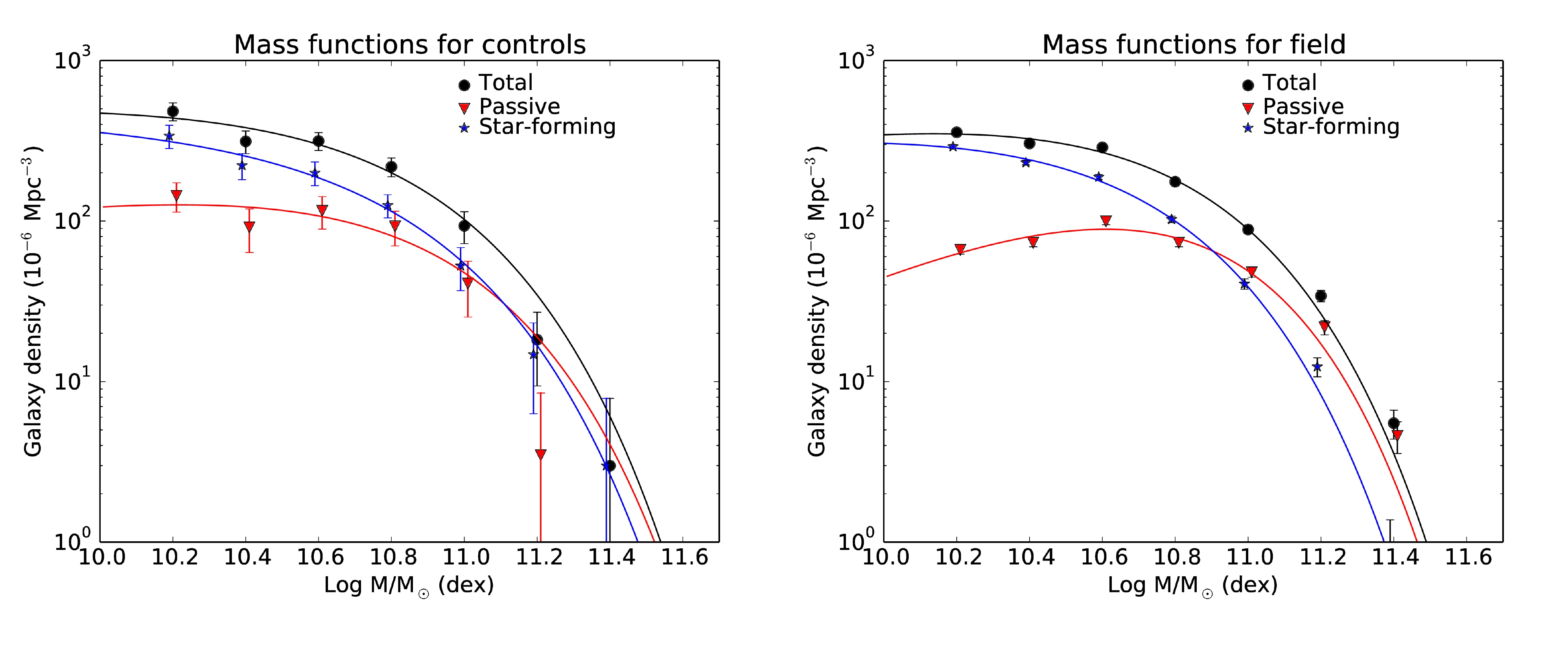}
\caption{Stellar mass functions for SpARCS foreground control galaxies (left) and UltraVISTA (right) field galaxies at 1.251 $<$ z $<$ 1.487.  Passive and star-forming stellar mass data are offset by $\pm$ 0.01 dex in stellar mass for clarity.}
\label{appendix_func}
\end{figure*}

Figure A1 shows the stellar mass functions and passive and star-forming fractions for simulation-averaged SpARCS foreground (controls) vs. UltraVISTA field galaxies.  All normalizations and parameters are within 2$\sigma$ of one another between the two field samples, and most are within 1$\sigma$, indicating overall similarity.  The highest-mass bins show the most difference from UltraVISTA, but these bins have very low absolute galaxy counts ($\sim$ 2 $\pm$ 1 and $\sim$ 0.3 $\pm$ 0.5 for the second-highest and highest mass bins).  The quenching efficiency of the SpARCS field compared to the UltraVISTA field is consistent with zero: 4 $\pm$ 4 \%.  All of these results indicate that there are no significant differences detectable between our SpARCS field galaxies and UltraVISTA field galaxies in terms of their stellar mass functions and passive fractions, particularly at low to intermediate masses, verifying the robustness of our principal results and the quality of our data.

\end{appendix}

\end{document}